\documentclass{article}
\author{Claire Levaillant}
\title{Making a circulant $2$-qubit entangling gate}
\usepackage{epsfig, amsmath, amsthm}
\newcommand{\nts}{\negthickspace}
\begin{document}
\maketitle
\begin{center}
\textbf{Abstract}
\end{center}
We present a way to physically realize a circulant $2$-qubit entangling gate in the Kauffman-Jones version of $SU(2)$ Chern-Simons theory at level $4$. Our approach uses qubit and qutrit ancillas, braids, fusions and interferometric measurements. Our qubit is formed by four anyons of topological charges $1221$. Among other $2$-qubit entangling gates we generate in the present work, we produce in particular the circulant gate
$$CEG=\frac{1}{4}\,I+i\frac{\sqrt{3}}{4}\,J-\frac{3}{4}\,J^2+i\frac{\sqrt{3}}{4}\,J^3$$ where $J$ denotes the permutation matrix associated with the cycle $(1432)$ and $I$ denotes the identity matrix.

\section{Introduction}
We present a way to produce a $2$-qubit entangling gate. With additional braiding, we obtain the circulant $2$-qubit entangling gate announced in the abstract.
Our core idea relies on the fact that if $G$ is a gate acting on the $2$-qubit $1221$ that is not entangling, then for each column $i$ of $G$, we must have
$$g_{11}^{(i)}g_{33}^{(i)}-g_{13}^{(i)}g_{31}^{(i)}=0$$
As soon as the equality above is not satisfied on a column $i$ of the gate, it is entangling. This will for instance be the case if the $2$-qubit $|11>$ gets mapped to a linear combination $\lambda\,|11>+\mu\,|33>$ with $\lambda$ and $\mu$ two non-zero complex scalars. We will make a gate of the form
$$\begin{array}{cc}&\begin{array}{cccc}\;\;|11>&|13>&|31>&\nts|33>\end{array}\\\begin{array}{l}|11>\\|13>\\|31>\\|33>\end{array}&\nts\begin{pmatrix}\star&&&0&&&0&&&\star \\
0&&&\star&&&\star&&&0\\0&&&\star&&&\star&&&0\\\star&&&0&&&0&&&\star\end{pmatrix}\end{array}$$
We proceed by "qubit transfer". We need two qubit ancillas $\frac{1}{\sqrt{2}}\,(|1>+|3>)$ which will ultimately carry the output. These ancillas are made in \cite{CL3} from the qubit $1122$ by braiding only. The main steps of the winning protocol (we call "winning protocol" a protocol in which all the measurements result in a desired outcome) are summarized below.
\begin{itemize}
\item
By interferometry and perhaps recovery, we fuse each ancilla into each qubit input. The process is described at length in \cite{CL4}.
\item We run an interferometry on $4$ anyons and hope to measure $0$. The $(0,0)$ and $(4,4)$ contributions (see forthcoming discussion) will force the output we like. Immediately after, we use a $qutrit$ ancilla $\frac{1}{\sqrt{2}}(|0>+|4>)$ and run a second interferometry aimed at getting rid of the $(2,2)$-contribution from previous measurement.
\item Get back to a $2$-qubit by additional fusions, braids and measurement.
\end{itemize}
First, we will go over the "winning protocol", then we will discuss the recovery procedures for undesired outcomes after each of the two interferometric measurements of the second point above. There are two kinds of recovery procedures, namely those which lead to the gate and those which lead to retrieving an intact input.
\newtheorem{Theorem}{Theorem}
\begin{Theorem}
The following protocol, with the measurement outcomes like indicated on the figure below
\begin{center}
\epsfig{file=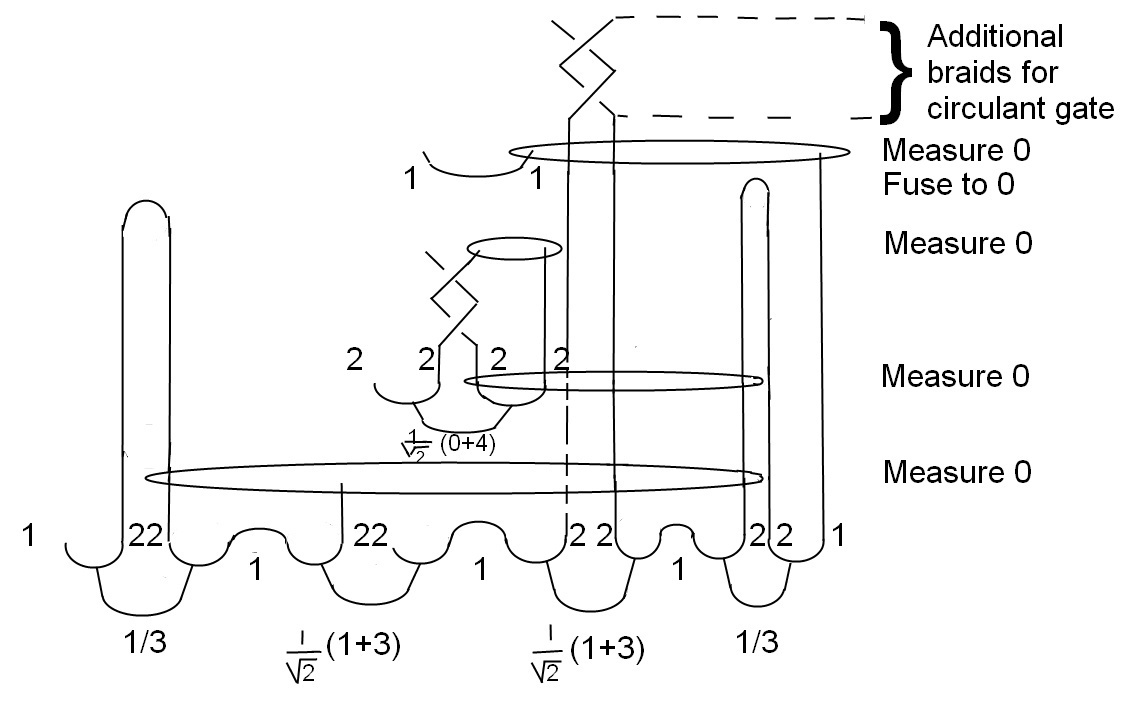, height=8.5cm}
\end{center}

produces the $2$-qubit entangling gate
$$\begin{array}{cc}&\begin{array}{cccc}|11>&|13>&|31>&|33>\end{array}\\\begin{array}{l}|11>\\\\|13>\\\\|31>\\\\|33>\end{array}&\begin{pmatrix}
-\frac{1}{2}&0&0&-i\frac{\sqrt{3}}{2}\\\\0&-\frac{1}{2}&-i\frac{\sqrt{3}}{2}&0\\\\
0&-i\frac{\sqrt{3}}{2}&-\frac{1}{2}&0\\\\-i\frac{\sqrt{3}}{2}&0&0&-\frac{1}{2}\end{pmatrix}\end{array}$$
If the measurement outcomes are not like on the figure, then by recovery we can still produce the gate given above. \\
If we do further a full twist like on the figure, we get the circulant gate $$C\bigg(\frac{1}{4},\frac{i\sqrt{3}}{4},-\frac{3}{4},\frac{i\sqrt{3}}{4}\bigg)$$ which we named CEG.

$$CEG= \begin{array}{cc}&\begin{array}{cccc}|11>&|13>&|31>&|33>\end{array}\\&\\\begin{array}{l}|11>\\\\|13>\\\\|31>\\\\|33>\end{array}&\begin{pmatrix}
\frac{1}{4}&&i\frac{\sqrt{3}}{4}&&-\frac{3}{4}&&i\frac{\sqrt{3}}{4}\\\\i\frac{\sqrt{3}}{4}&&\frac{1}{4}&&i\frac{\sqrt{3}}{4}&&-\frac{3}{4}\\\\
-\frac{3}{4}&&i\frac{\sqrt{3}}{4}&&\frac{1}{4}&&i\frac{\sqrt{3}}{4}\\\\i\frac{\sqrt{3}}{4}&&-\frac{3}{4}&&i\frac{\sqrt{3}}{4}&&\frac{1}{4}\end{pmatrix}\end{array}$$
On the figure, the interrupted dots signify that the corresponding particle is not part of the interferometric measurement.
\end{Theorem}
\section{Proof of the Theorem}
\subsection{Setting}
The anyonic system we work with is the Kauffman-Jones version of $SU(2)$ Chern-Simons theory at level $4$. There are five particle types of topological charges $0$, $1$, $2$, $3$, $4$ and of respective quantum dimensions $1$, $\sqrt{3}$, $2$, $\sqrt{3}$, $1$. They obey to fusion rules like in \cite{KL}. The value of the Kauffman constant is $A=i\,e^{-i\frac{\pi}{12}}$. We use a unitary version of the theory as explained in \cite{ZW}. That is we use unitary $6j$-symbols and unitary theta symbols. We assume the reader is familiar with the basics of recoupling theory whose nice exposition can be found in \cite{KL}.
\subsection{Preliminary remark}
Before we start the proof, it will be useful to recall some axiomatics with trivalent vertices of Jones-Wenzl projectors, see $\S\,9.16$ of \cite{KL}.
\begin{center}
\epsfig{file=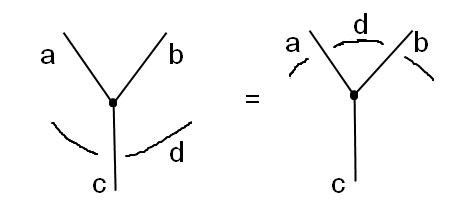, height=4cm}
\end{center}
For instance, in some situations, a charge line in the back can be moved by braiding at no cost. This is illustrated below. 
\begin{center}
\epsfig{file=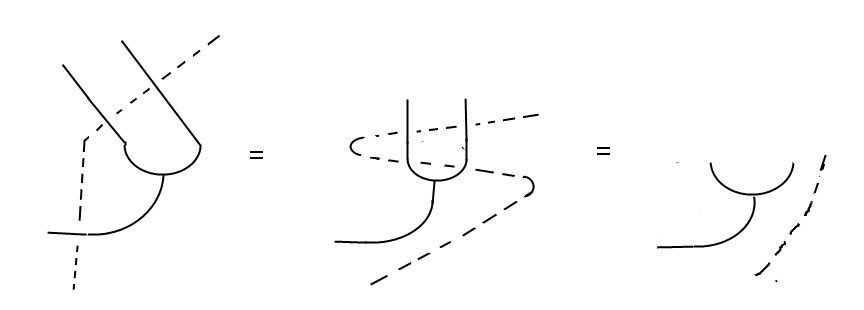, height=4.5cm}
\end{center}
\noindent The first equality holds by the move from the axiomatics and the second equality holds by applying a Reidemeister move twice. 
Along the proof, we will use such moves quite freely and without mentioning them. The reader should keep track of all the relative positions of the anyons on the figures, where the different charge lines go and how the strands get moved at no cost.

\subsection{With the measurement outcomes like on the figure}
We first show the Theorem in the case when all the measurement outcomes are like on the figure. We leave for later the exposition of the recovery procedures when the measurement outcomes are not the desired ones. We will assume the reader is familiar with the concept of interferometric measurement and draw the picture corresponding to after the two interferometric measurements of the protocol.
\begin{center}
\epsfig{file=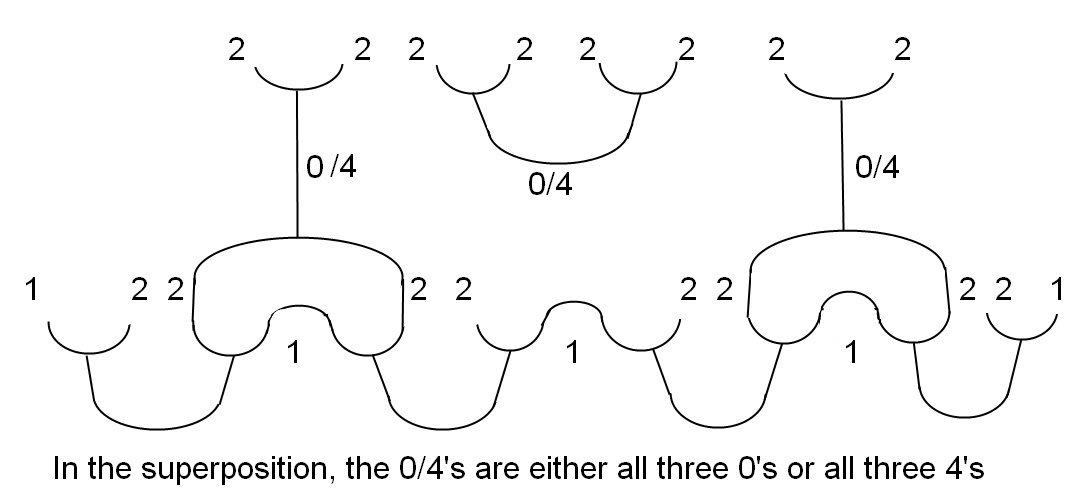, height=6cm}
\end{center}
The following steps present no difficulties. It remains to dispose of the qutrit ancilla. To do so, we do a full twist. We recall below the matrix for a full $\sigma_2$ twist on the qutrit $2222$. It is given by
$$\sigma_2^2(2,2,2,2)=\begin{pmatrix}-\frac{1}{2}&0&-i\frac{\sqrt{3}}{2}\\
0&-e^{i\frac{\pi}{3}}&0\\
-i\frac{\sqrt{3}}{2}&0&-\frac{1}{2}\end{pmatrix}$$
Hence, if after the full twist, we measure $0$ like on the figure of the Theorem, then we obtain the gate announced after we have split again the two qubits by a usual process of bringing a pair out of the vacuum and running an interferometric measurement. \\
Finally, if we do the two additional braids, we act on the right qubit by $\sigma_2^2(1,2,2,1)$. Hence we must multiply the $4$ blocks of the entangling gate to the left by
$$\sigma_2^2(1,2,2,1)=\begin{pmatrix}-\frac{1}{2}&&-i\frac{\sqrt{3}}{2}\\&&\\-i\frac{\sqrt{3}}{2}&&-\frac{1}{2}\end{pmatrix}$$
It yields the circulant entangling gate $CEG$.
\subsection{Other measurement outcomes}
\begin{itemize}
\item As far as the interferometric measurement aimed at the separation of both qubits, if we measure $2$ instead of $0$, we simply braid and remeasure. \\
\item Regarding the fusion measurements. If the two anyons of topological charge $2$ fuse into either $2$ or $4$, then we do one more fusion whose outcome will be necessarily $2$ by the fusion rules. Below, we made a zoom on the left qubit side. We do the same thing on the right qubit side.

\begin{center}
\epsfig{file=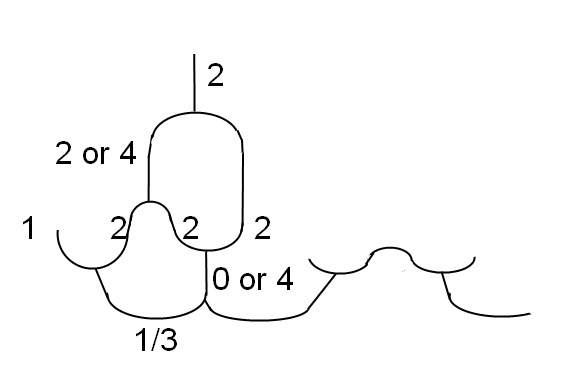, height=7cm}
\end{center}

When the first fusion outcome is $2$, we must keep track of a relative minus sign between the $0$ and the $4$ contribution. So, \\\\
\textit{If the first fusion outcome is $2$ \textbf{on one side and one side only}, all the anti-diagonal coefficients of the gate get multiplied by $-1$. Consequently, we get the conjugate transpose matrix, that is the inverse gate. Run the same protocol again as many times as is necessary to eventually produce the gate.}

\item If after the full twist, we measure $4$ instead of $0$, the gate formed has coefficients $-i\frac{\sqrt{3}}{2}$ on its diagonal and coefficients $-\frac{1}{2}$ on its anti-diagonal. In order to swap both diagonals and form the gate we want, it will suffice to do a Freedman type fusion operation on both qubits (left and right).

\begin{center}
\epsfig{file=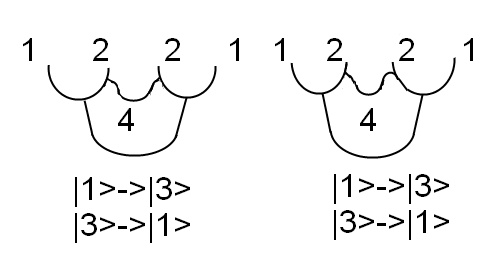, height=6cm}
\end{center}

\item Suppose the interferometric measurement involving the qutrit $2222$ has outcome $4$ instead of $0$. The only difference with before is
that when we deal with the $(0,0)$ (resp $(4,4)$) contribution, the qutrit before undergoing the full twist followed by the measurement carries the charge $4$ (resp $0$). Now we see how to conclude.\\\\
\textit{Apply the Freedman fusion operation on each side \textbf{only when} the two measurement outcomes involving the qutrit are \textbf{distinct}}.\\

\item Suppose the interferometric measurement involving the qutrit $2222$ has outcome $2$ instead of $0$.
Then the $(0,0)$ and $(4,4)$ contributions vanish and we get the picture
\begin{center}
\epsfig{file=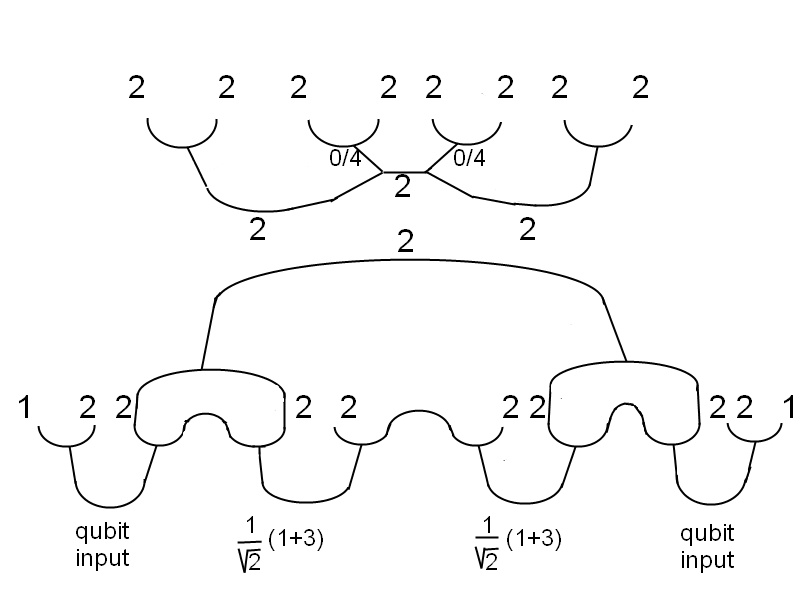, height=8.5cm}
\end{center}

The next picture summarizes what we do next. \\
\begin{itemize}
\item The two bubbles can be collapsed \textbf{at the cost of some minus signs} when the qubit input and qubit ancilla carry distinct charges.
\item We move one step forward towards the recovery by braiding like on the new figure below.
\item The upper component is independent from the input, hence can be disposed of (for instance by fusion) after the braiding has occurred.
\end{itemize}
On the figure below, we summarize the three steps that have just been described.

\begin{center}
    \epsfig{file=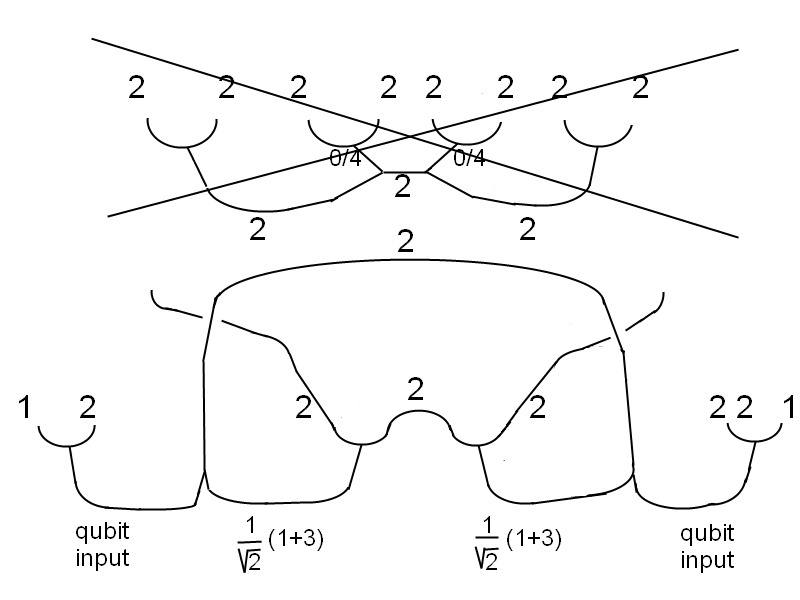, height=7cm}
    \end{center}

Next, if we undo the two braids by F-move and R-move, we get a superposition
\begin{center}
\epsfig{file=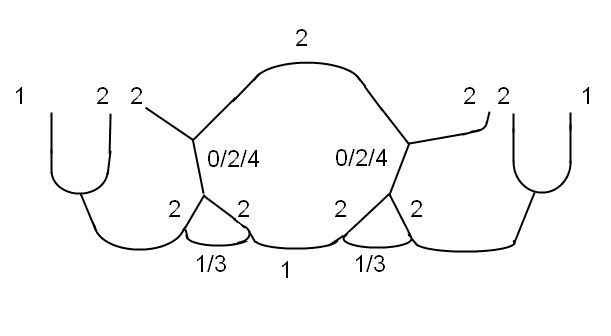, height=5.5cm}
\end{center}
\textbf{We get here to one of the prettiest features of the protocol}. Namely, we recall from previous step that when the qubit input and the qubit ancilla carry distinct charges, the contributions are opposite from when they carry identical charges. When the charge line from the F-move on the figure above carries the label $0$ or $4$ however, the two F-symbols involved in the triangular bubble collapse are identical for a fixed input $|1>$ or $|3>$. Thus, the contributions in $0$ and $4$ are null. On the other hand, when the charge line from the F-move on the figure carries a label $2$, the two F-symbols involved in the same collapse are opposite for a fixed input $|1>$ or $|3>$. Again, when the qubit input and the qubit ancilla carry distinct charges, get a relative minus sign and so globally there is no relative minus sign to keep track of depending on what the qubit input is. Hence the new diagram.
\begin{center}
\epsfig{file=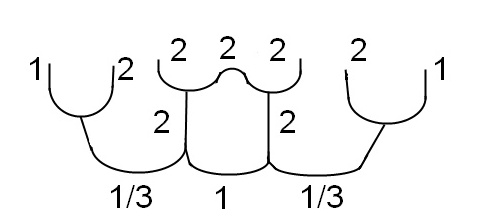, height=6cm}
\end{center}
  Now we measure the last three anyons.
 \begin{center}
 \epsfig{file=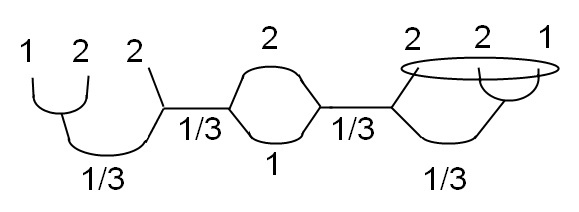, height=4cm}
 \end{center}
 Whether we measure a $1$ or a $3$, the measurement introduces a minus sign when the two qubit inputs are distinct.  The next steps are
 \begin{itemize}
 \item Bring a pair of $1$'s (or $3$'s depending on last measurement's outcome) out of the vacuum and run an interferometry on the four anyons to the right. Hope to measure $0$ and if don't, braid and remeasure until measure $0$.
 \item In the case when the measurement outcome was $3$, do a $4$ pair fusion on each qubit to get it back to the initial shape $1221$.
 \item Correct the minus signs by doing a full $\sigma_1$ twist on each qubit input in order to get back to an \textbf{intact} $2$-qubit input. This closes the recovery.
 \item Run the main protocol all over again with the intact $2$-qubit input.
 \end{itemize}

We must still discuss the cases when the core interferometric measurement yield outcomes $2$ or $4$.

\item Suppose the main interferometric measurement outcome is $4$.
If by successive F-move, we build the tree superposition arising from the interferometric measurement, all the coefficients $\lambda_{ij}$'s of the superposition are equal.
\begin{center}
\epsfig{file=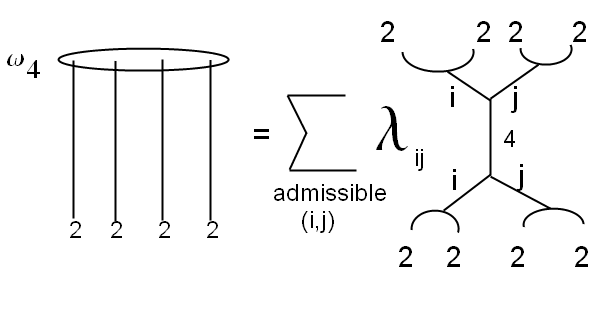, height=6cm}
\end{center}
The pair $\lbrace i,j\rbrace$ must be $\lbrace 2,2\rbrace$ or $\lbrace 0,4\rbrace$. 
We use the same qutrit ancilla as before. We get the following diagram.
\begin{center}
\hspace{-0.3cm}\epsfig{file=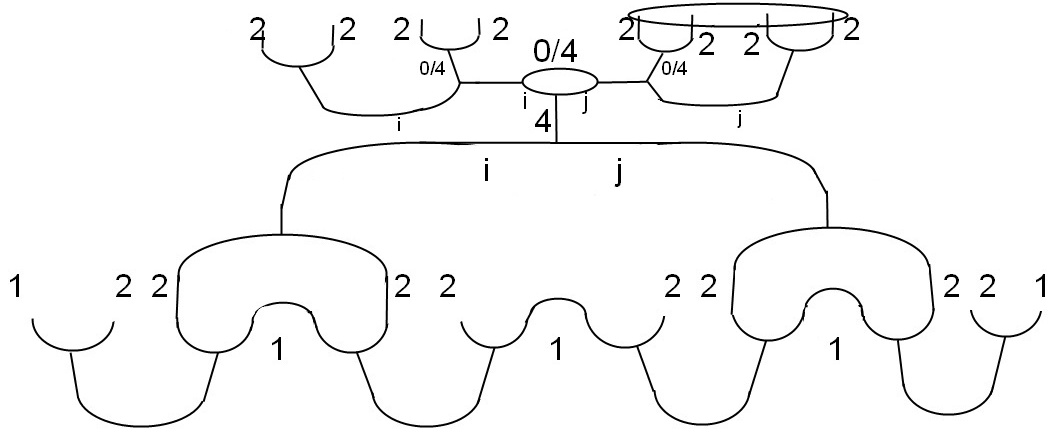, height=4.5cm}
\end{center}
The two unlabeled edges at the top of the figure are those arising from the two F-moves. All the $6j$-symbols involved are equal to $1$. Because the qutrit ancilla used is a superposition of charges $0$ and $4$, when $j=0$ or $j=4$, the unlabeled edge from the F-move is also such a superposition. By running an interferometric measurement on the $4$ anyons like indicated on the figure, we force the value to either $0$ or $4$ and the vertical mirror edge to the other value by the fusion rules. This holds unless the outcome from the measurement is $2$, in which case the contribution that is left is the one arising from the values $i=j=2$. Suppose first we measure $0$ or $4$.
\begin{itemize}
\item If we measure $0$, we get the superposition
\begin{center}
\epsfig{file=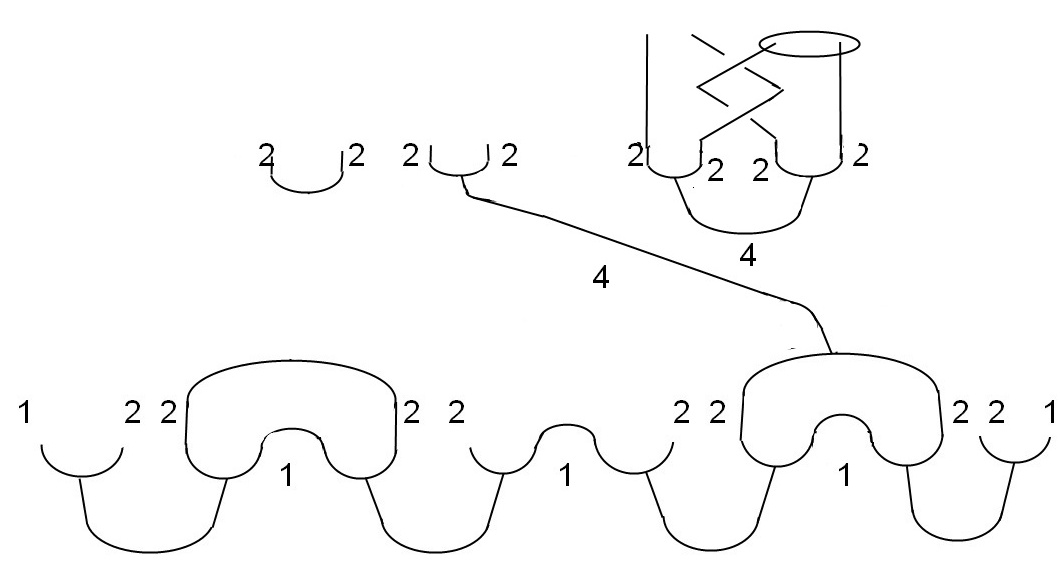, height=6cm}
\end{center}
\begin{center}
\epsfig{file=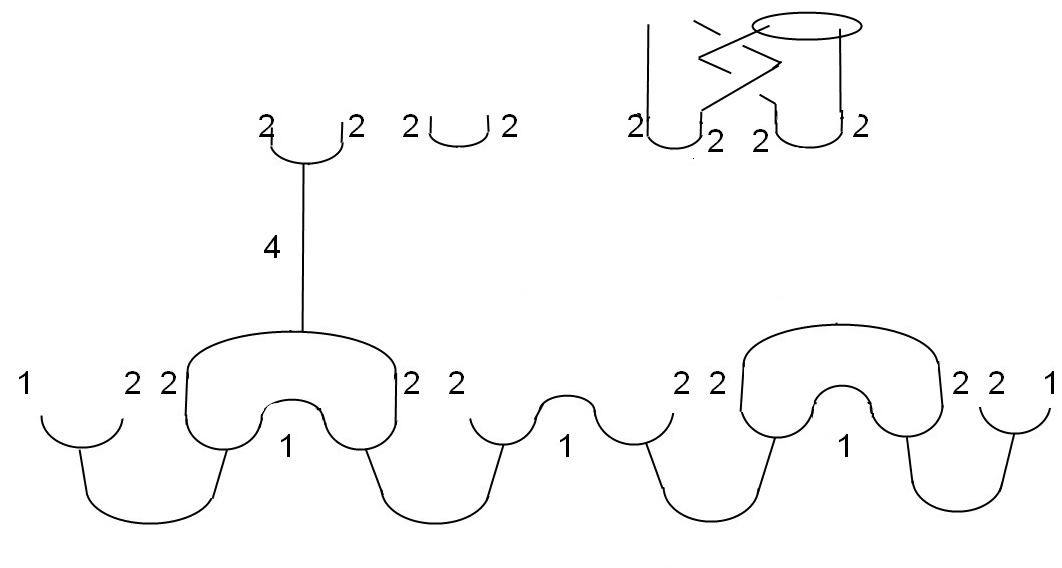, height=6cm}
\end{center}
\item If we measure $4$, we get the superposition
\begin{center}
\epsfig{file=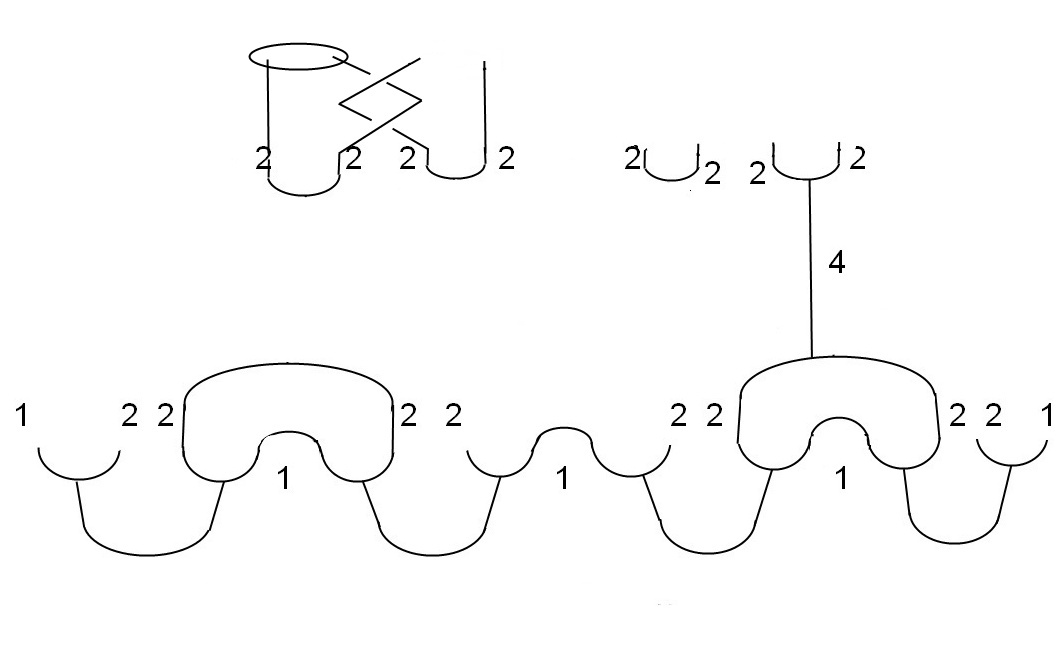, height=6.5cm}
\end{center}
\begin{center}
\epsfig{file=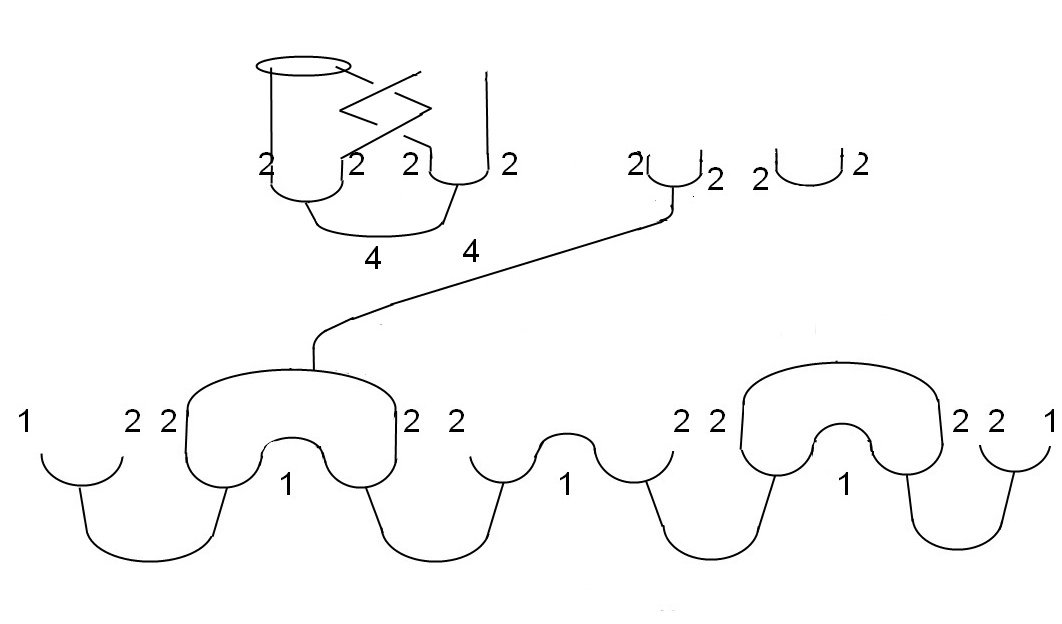, height=6.5cm}
\end{center}

\end{itemize}
By symmetry, the second case can be deduced from the first case, so we first deal with the $0$ outcome.
The first step is to remove the last four anyons at the top by braiding and measurement, like done in the past.
After that, we may indeed remove them at some cost and to finish, we fuse anyon number $2$ (resp number $5$) "at the bottom" with anyon number $1$ (resp number $4$) "at the top". With the same consequences as already discussed in the past, depending on the fusion outcomes. The last step is the usual separation step (bring a pair of $1$'s out of the vacuum, measure the last four anyons by interferometry, braid the central anyons and remeasure until the outcome is $0$). We are ready to give the shape of the gate. Suppose the outcome of the measurement happening after the full twist is $0$. Then, we get the gate
$$\begin{array}{l}\end{array}$$
$$\begin{array}{cc}&\begin{array}{cccc}|11>&|13>&|31>&|33>\end{array}\\&\\\begin{array}{l}|11>\\\\|13>\\\\|31>\\\\|33>\end{array}&\begin{pmatrix}
0&-i\frac{\sqrt{3}}{2}&-\frac{1}{2}&0\\&&&\\-i\frac{\sqrt{3}}{2}&0&0&-\frac{1}{2}\\&&&\\-\frac{1}{2}&0&0&-i\frac{\sqrt{3}}{2}\\&&&\\
0&-\frac{1}{2}&-i\frac{\sqrt{3}}{2}&0\end{pmatrix}\end{array}$$
$$\begin{array}{l}\end{array}$$
In light of the discussion we had before, the gate we get is different only when \underline{\underline{\textbf{exactly one}}} of the two multi-channel fusions outcomes is $2$. In the latter case, we must replace all the $-\frac{1}{2}$ with $\frac{1}{2}$ or indifferently all the $-i\frac{\sqrt{3}}{2}$ with $i\frac{\sqrt{3}}{2}$ (up to a $\pi$ overall phase, it is the same gate). \\\\
If the outcome of the measurement happening after the full twist is $4$, swap the positions of the $-\frac{1}{2}$ and $-i\frac{\sqrt{3}}{2}$ coefficients. \\\\
In order to always get the same gate like arbitrarily decided, here is how we proceed depending on what we measure.
\begin{itemize}
\item \underline{\textbf{Independently from the fusions outcomes}}, if measure $0$ (resp $4$) after the interferometric measurement that follows the full twist, then do a Freedman type fusion operation on the left (resp right) qubit.
\item If \underline{\underline{\textbf{exactly one}}} of the two multi-channel fusions outcomes is $2$, then get the inverse gate and start all over again.
\end{itemize}
If the first interferometric measurement involving the qutrit ancilla has outcome $4$ instead of $0$ (see second series of figures above), then keep the same protocol, except swap the two words "left" and "right". \\\\
It remains to study the case when the first interferometric measurement involving the qutrit ancilla has outcome $2$ instead of $0$ or $4$. In that case, it forces $i=j=2$. We get after simplifications, braids and fusions,
\begin{center}
\epsfig{file=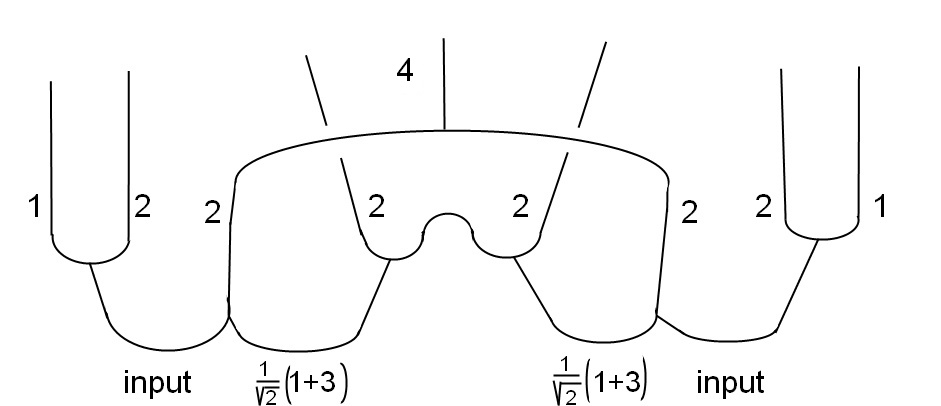, height=6cm}
\end{center}
By similar arguments as before, the diagram reduces to
\begin{center}
\epsfig{file=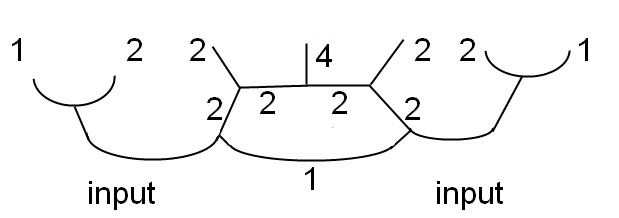, height=4cm}
\end{center}
And after F-moves, we obtain
\begin{center}
\epsfig{file=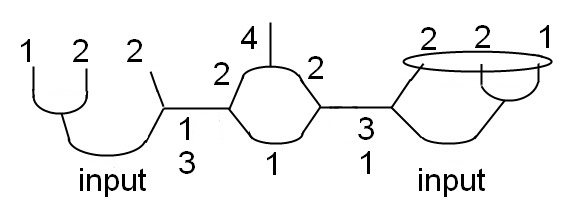, height=5cm}
\end{center}
Then, measure the last three anyons to the right like on the figure. Projecting the horizontal edge onto $1$ (resp $3$) also forces the label of the vertical mirror edge, that is $3$ (resp $1$). Without loss of generality, suppose we measure $3$. Then, we get the picture (after collapsing the bubble, bringing pairs out of the vacuum and measuring by interferometry on both sides,
\begin{center}
\epsfig{file=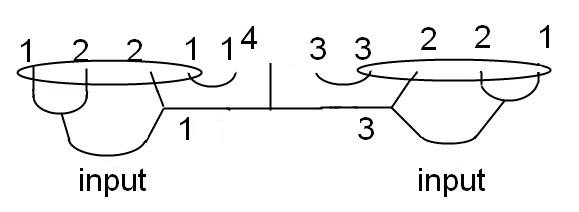, height=5cm}
\end{center}
And to finish,
\begin{center}
\epsfig{file=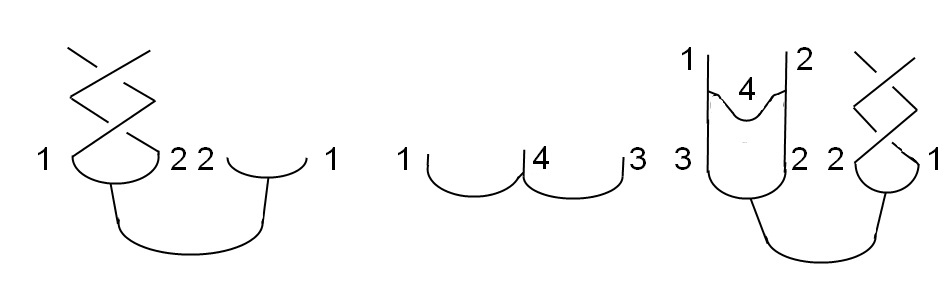, height=4cm}
\end{center}
Like explained before, the full twists are there to correct the minus signs. The central "omega" can be disposed of. We retrieve an intact $2$-qubit input ready to be acted on again. End of the recovery in that case.
\item The last case to study is when the main interferometric measurement is $2$. Here, we will do a complete recovery, that is one at the end of  which we retrieve an intact input that is ready to be acted on again. We have the picture
\begin{center}
\epsfig{file=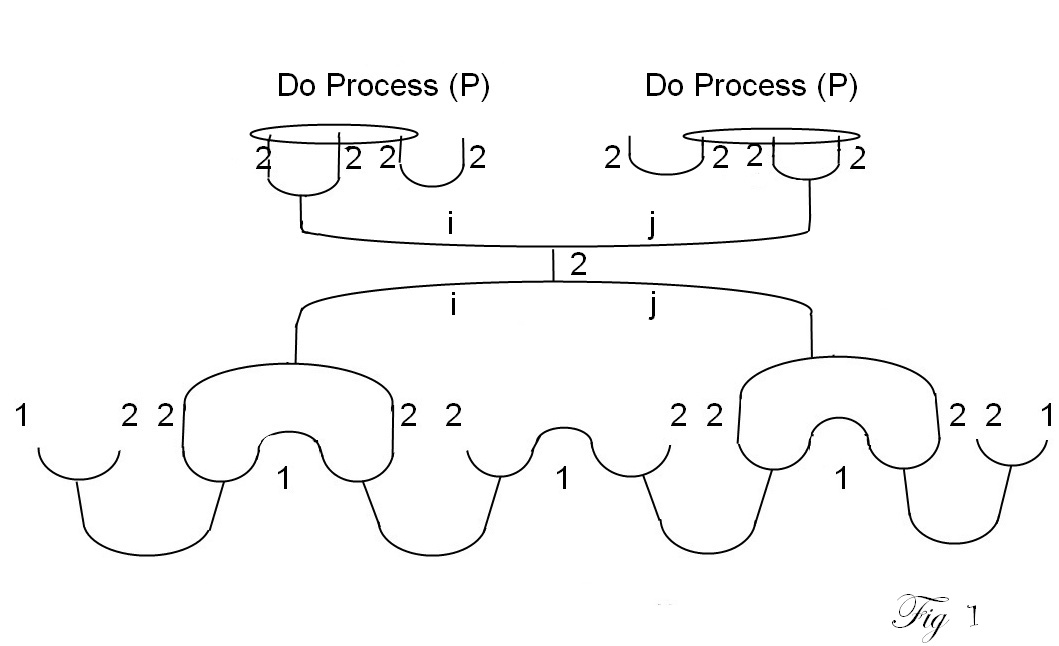, height=7cm}
\end{center}
The diagram above is a superposition of all possible labelings $(i,j)$. One of $i$ or $j$ can be $2$ and the other one $0$ or $4$. Or both $i$ and $j$ are $2$. \\
We will describe a process which allows to either project a qutrit superposition $0/2/4$ onto the trit $2$ or onto the superposition $0/4$.
\newtheorem{Lemma}{Lemma}
\begin{Lemma} \textit{Process (P).} Project a qutrit superposition $0/2/4$ onto either $2$ or $0/4$.
\begin{center}
\epsfig{file=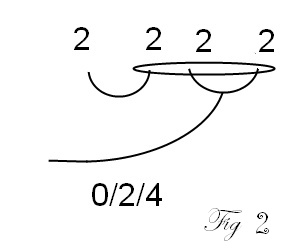, height=5cm}
\end{center}
If the outcome of the interferometric measurement of the three anyons of the figure is either $0$ or $4$, we have projected onto $|2>$, end of the process.\\
If the outcome is rather $2$, then we braid like on the figure below.
\begin{center}
\epsfig{file=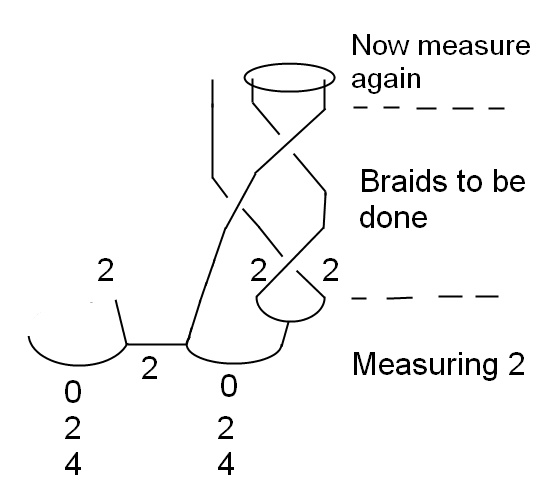, height=8cm}
\end{center}
If the measurement from the top has outcome $0$, we remove the pair of $2$'s and start the process again with an intact qutrit $0/2/4$. \\
If the measurement outcome is $4$, we fuse the last three anyons two by two from the right, then fuse a pair of $4$'s into the last two anyons, then start the process all over again. \\
If the measurement outcome is $2$, fuse the last three anyons two by two from the right, then braid the last two anyons. As a result of the process, we have projected the qutrit onto a superposition $0/4$.
\end{Lemma}

\textsc{Proof of Lemma}. When the outcome is $2$, we get a superposition $0/2/4$ with the amplitude in $|2>$ multiplied by a factor $\frac{1}{\sqrt{2}}$ (arising from the F-move on the zero labelled vertical edge between the ancilla pair of $2$ and the qutrit). Up to overall phase the matrix of the braiding action on the qutrit $2222$ from the figure is given by
$$\begin{array}{cc}&\begin{array}{ccc}|0>&|2>&|4>\end{array}\\&\\\begin{array}{l}|0>\\\\|2>\\\\|4>\end{array}&\begin{pmatrix} \frac{1}{2}&\frac{1}{\sqrt{2}}&\frac{1}{2}\\&&\\\frac{1}{\sqrt{2}}&0&-\frac{1}{\sqrt{2}}\\&&\\\frac{1}{2}&-\frac{1}{\sqrt{2}}&\frac{1}{2}\end{pmatrix}
\end{array}$$
Suppose the second measurement outcome is $0$. After removing the pair of $2$'s, we are back to the initial configuration of Fig. $2$ and we start the process all over again. \\
Suppose the second measurement outcome is $4$. Fuse the last three anyons to the right two by two. Then fuse a pair of $4$'s into the last two anyons in order to map the $|2>$ qutrit into its opposite. Then, we are back to the initial configuration of Fig. $2$ and ready to start the process all over again.\\
Suppose the second measurement outcome is $2$. After the two fusions, the effect of the $\sigma_1$-braid is simply to correct the $\pi$ relative phase between the $|0>$ trit and the $|4>$ trit that was introduced after the series of braids and the final measurement. This finishes the proof of Lemma $1$.


Suppose we measure $0$ or $4$ during the process $(P)$. Then $i=j=2$. Braid and fuse in order to get to the following configuration
\begin{center}
\epsfig{file=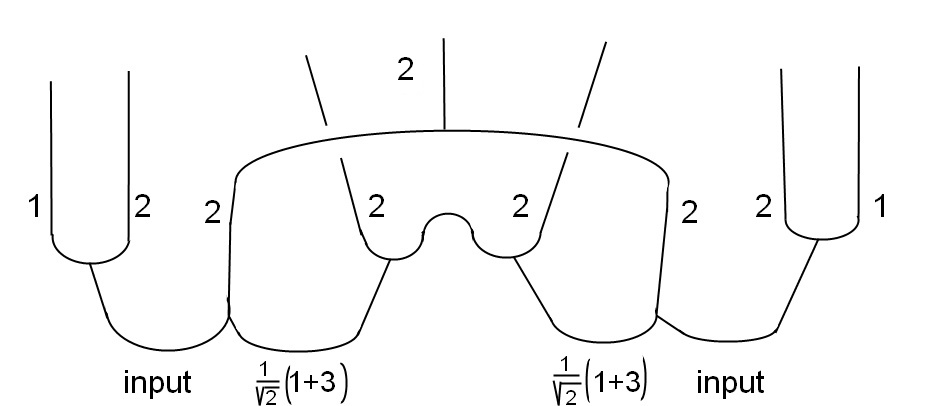, height=5.5cm}
\end{center}
By similar arguments as before, the diagram reduces to
\begin{center}
\epsfig{file=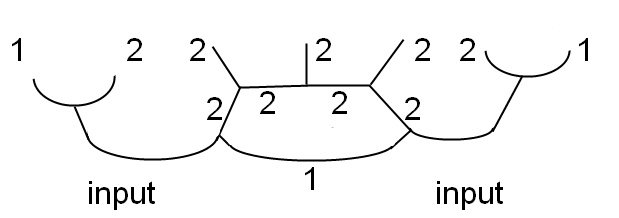, height=5cm}
\end{center}
After two F-moves, we get
\begin{center}
\epsfig{file=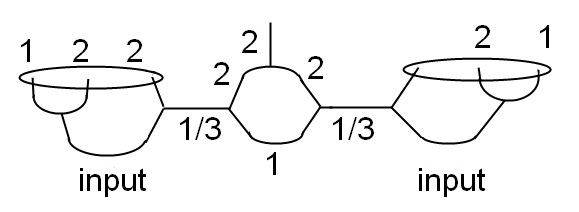, height=5cm}
\end{center}
Like on the figure, we run the two indicated interferometric measurements.
\begin{itemize}
\item If the outcomes are identical on both sides, we have a relative minus sign when the two qubit inputs are distinct.
\item If the outcomes are distinct on the two sides, we have a relative minus sign when the two qubit inputs are identical.

\end{itemize}
Thus, since we work up to overall phase, it does not matter what we measure, all what matters are the measurements themselves.
Suppose in most generality we measure $1$ to the left hand side and $3$ to the right hand side. Then place a pair of $1$'s to the left hand side and a pair of $3$'s to the right hand side. Then measure by interferometry the four anyons to the extreme left and the four anyons to the extreme right.
\begin{center}
\epsfig{file=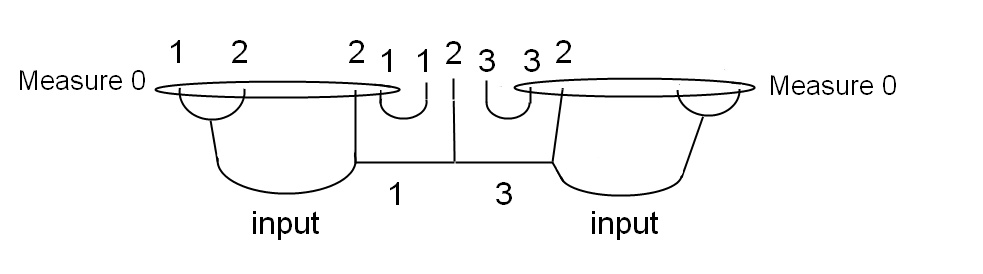, height=4cm}
\end{center}
And to finish,
\begin{center}
\epsfig{file=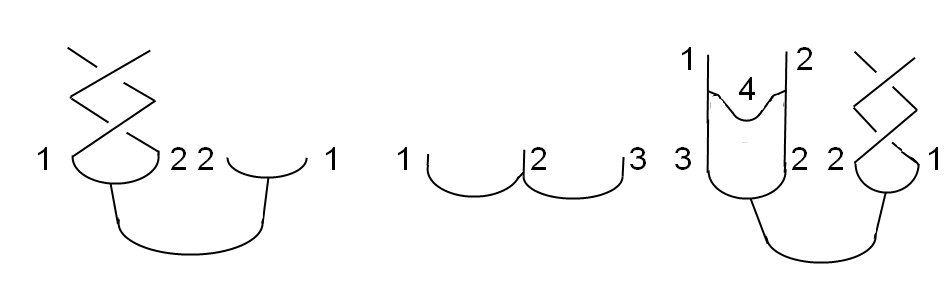, height=4cm}
\end{center}
We can dispose of the "central omega" and after doing so retrieve an intact input. End of the recovery in that case.

Suppose we measure $2$ the first time we use Process (P), say without loss of generality to the left hand side of Fig. $1$. Then, by the process, we are able to project onto $i=0,4$. It then forces $j=2$. We get the picture
\begin{center}
\epsfig{file=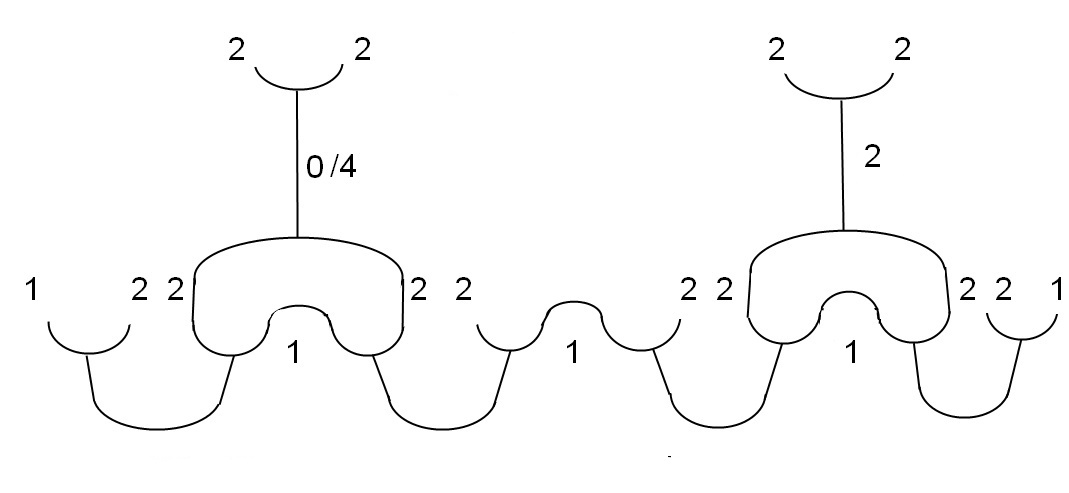, height=5cm}
\end{center}

After collapsing the bubbles, doing F-moves, we get the new picture, where the interferometric measurements are used to project the related horizontal edges onto either $1$ or $3$. 

\begin{center}
\epsfig{file=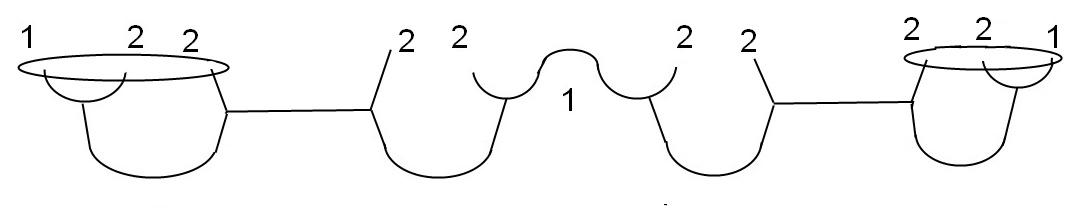, height=2.5cm}
\end{center}
Whenever the measurement outcome is $3$ instead of $1$, fuse a pair of $4$'s into the $2$'s in order to switch the $3$ label to a $1$ label. 
Start the protocol over again from the configuration
\begin{center}
\epsfig{file=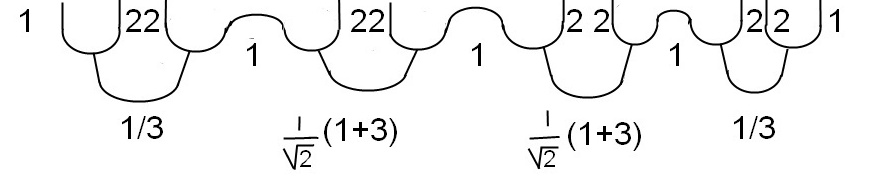, height=2.5cm}
\end{center}
\end{itemize}

\noindent \textbf{Acknowledgements.} The author is very happy to thank Mike Freedman for helpful discussions. She thanks Kaushal Patel and Brayden Ware for their kind support.

\end{document}